\documentclass[iop]{emulateapj}
\usepackage{natbib}
\usepackage[backref,breaklinks,colorlinks,citecolor=blue]{hyperref}        
\usepackage[all]{hypcap}
\usepackage{amsmath}
\usepackage{apjfonts}

\shorttitle{Silicate Composition of the ISM}
\shortauthors{S. Fogerty et al.}

\begin{document}

\title{Silicate Composition of the Interstellar Medium}

\author{
S.     Fogerty\altaffilmark{1}, 
W.     Forrest\altaffilmark{1}, 
D. M.  Watson\altaffilmark{1}, 
B. A.  Sargent\altaffilmark{2}, and 
I.     Koch\altaffilmark{1,3}}


\affil{$^1$Department of Physics and Astronomy, University of Rochester, Rochester, NY 14627, USA; \href{mailto:sfogerty@pas.rochester.edu}{sfogerty@pas.rochester.edu}\\
$^2$Center for Imaging Science and Laboratory for Multiwavelength Astrophysics, Rochester Institute of Technology\\
54 Lomb Memorial Drive, Rochester, NY 14623, USA\\
$^3$Department of Earth \& Planetary Sciences, Washington University, St. Louis, MO 63130, USA}



\begin{abstract}
The composition of silicate dust in the diffuse interstellar medium and in protoplanetary disks around young stars informs our understanding of the processing and evolution of the dust grains leading up to planet formation.  Analysis of the well-known 9.7$\mu$m feature indicates that small amorphous silicate grains represent a significant fraction of interstellar dust and are also major components of protoplanetary disks. However, this feature is typically modelled assuming amorphous silicate dust of olivine and pyroxene stoichiometries. Here, we analyze interstellar dust with models of silicate dust that include non-stoichiometric amorphous silicate grains. Modelling the optical depth along lines of sight toward the extinguished objects Cyg OB2 No. 12 and $\zeta$ Ophiuchi, we find evidence for interstellar amorphous silicate dust with stoichiometry intermediate between olivine and pyroxene, which we simply refer to as ``polivene.'' Finally, we compare these results to models of silicate emission from the Trapezium and protoplanetary disks in Taurus.
\end{abstract}


\keywords{(ISM:) dust, extinction --- infrared: stars --- protoplanetary disks --- stars: pre-main sequence --- stars: variables: T Tauri, Herbig Ae/Be}

\section{Introduction}
Studies of the dust in the interstellar medium (ISM) and in protoplanetary disks informs studies of the evolution of protoplanetary disks toward planetary systems. The dust in the ISM becomes the raw material of protoplanetary disk solids, which undergo processing during accretion disk evolution. In this way, the dust composition of the ISM and protoplanetary disks probe key phases leading up to planet formation. T Tauri star (TTS) disks show spectral features of dust in emission at infrared wavelengths observable by the Infrared Spectrograph (IRS; \citealp{houck04}) onboard the \textit{Spitzer Space Telescope} \citep{werner04}. Some studies (e.g. \citealt{prz03}; \citealt{kessler05,kessler06}; \citealt{watson09}) have characterized disks' dust content using spectral diagnostics such as spectral indices, which are measurements of flux ratios between wavelengths chosen to focus on the ``continuum'' between strong silicate features.  Studies such as that by \cite{sargent06,sargent09} characterize the dust in these disks through modelling disk emission in IRS spectra using theoretical and laboratory measured dust grain optical properties. With this approach, it is very important to ensure that the dust opacities used in the model are not only precise, but also representative of the actual dust in protoplanetary disk environments.

Silicate dust is of great interest to astronomers due to its prevalence in many different astrophysical environments, including the ISM, protoplanetary disks around young stars (e.g., TTSs, Herbig Ae/Be stars), evolved stars (e.g., asymptotic giant branch stars, red supergiant stars), massive stars (e.g., supergiant B[e] stars), and even the immediate environs of active galactic nuclei (ULIRGS and quasars). The broad $\sim$10$\mu$m and $\sim$20$\mu$m resonances of silicate dust cause increased extinction in the ISM near these wavelengths. In protoplanetary disks, the silicate dust in the relatively warmer outer layer of the disk gives rise to spectral emission features. While the silicate dust in the ISM is almost entirely in amorphous forms (see \citealt{mathis90}, \citealt{lidraine01}, \citealt{kemper04,kemper05}), the spectra of protoplanetary disks and Herbig AeBe stars also show evidence of more sharp, narrow emission features (henceforth ``fine structure''), indicating crystalline silicate dust in addition to amorphous silicates (see \citealt{bouwman01}, \citealt{forrest04}, \citealt{vanboekel04}, \citealt{sargent06, sargent09}, \citealt{sicilia07}, \citealt{watson09}). The smooth infrared spectral features near 10$\mu$m and 20$\mu$m indicate the presence of amorphous silicate dust, typically modelled as silicate grains of olivine and pyroxene stoichiometries. In their modelling, \cite{sargent06, sargent09} include three crystalline silicate dust components as well as amorphous silicates of both pyroxene and olivine stoichiometry. In these models, each dust component must be carefully selected, since silicate dust grains can come in a wide variety of size and shape distributions.

\begin{figure*}
\includegraphics[angle=90, width=\textwidth]{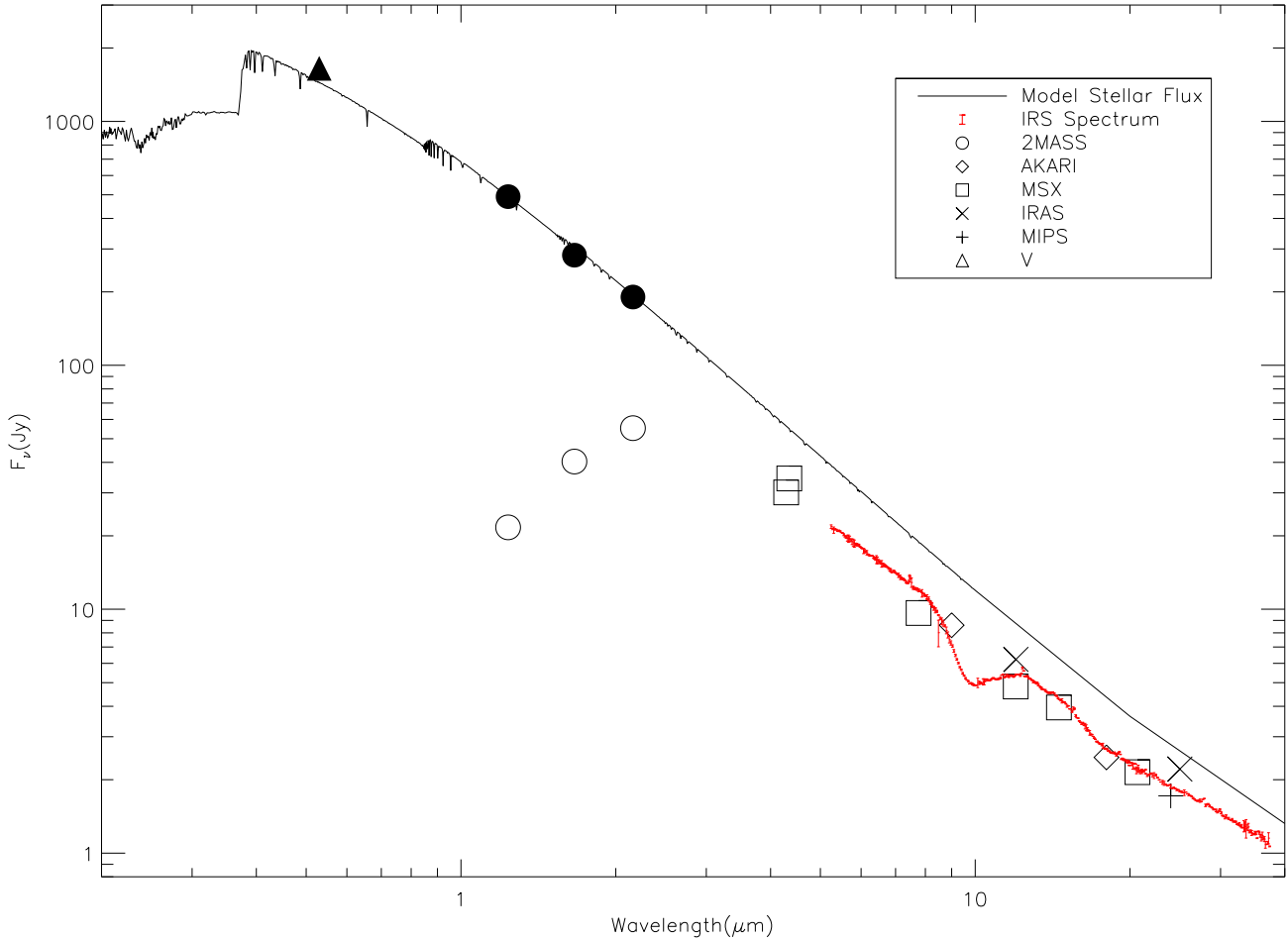}
\caption{Spectral Energy Distribution for Cyg OB2 No. 12. The Spitzer IRS spectrum is shown in red. The solid line represents the stellar flux, using a Kurucz stellar atmosphere model given the spectral type from \citealt{humphreys78}. A free-free emission component (corresponding to a wind) proportional to $\nu^{0.6}$ was added for wavelengths $>$1$\mu$m so that the free-free component is 10$\%$ of the stellar flux at 8$\mu$m. Solid symbols denote photometry corrected for extinction, while open symbols have no correction applied. Photometry from  \citealt{caballero14} (V, triangle), 2MASS (J, H, K$_s$; circle), AKARI (9 and 18$\mu$m; diamond), MSX (square), IRAS (12 and 25$\mu$m; x), and MIPS (24$\mu$m; plus) are shown.\label{fig1}}
\end{figure*}

While amorphous silicate dust grains in astrophysical contexts are typically modelled assuming a pyroxene or olivine stoichiometry, presolar dust grains embedded in primitive meteorites suggest that this dust is often non-stoichiometric. Models of the nature and composition of dust in the ISM can be informed by presolar grains found in meteorites and interplanetary dust particles. Analysis of the oxygen isotopic compositions of these presolar grains shows that they were formed in supernovae, red giant branch stars, and asymptotic giant branch stars (\citealt{choi98}, \citealt{nittler97b}, \citealt{nittler98}). The presolar grains traversed the ISM, eventually becoming a part of the forming solar system, where some were preserved in meteorites. Studies have found presolar silicate grains with stoichiometry intermediate between olivine and pyroxene (\citealt{boseThesis}, \citealt{bose12}, \citealt{floss09}, \citealt{stadermann09}, \citealt{vollmer09}). Models using interstellar and protoplanetary dust should consider amorphous grains of intermediate stoichiometry between olivine and pyroxene, given the meteoritic evidence for non-stoichiometric amorphous silicates from presolar grains.

Studies using the spectra of the dusty emission surrounding the Trapezium, as well as the extinction profile toward objects like Cyg OB2 No. 12 and $\zeta$ Oph, inform our understanding of the composition of interstellar dust. Work by \cite{gillett69} and \cite{gillett73} showed the Trapezium spectrum at near-infrared wavelengths was consistent with emission from silicate dust grains. \cite{draine84} based their derivation of ``astronomical silicate'' optical properties in part on the 8-13$\mu$m Trapezium spectrum. Cyg OB2 No. 12, a very luminous B5 hypergiant star at roughly ~1.7 kpc distance (\citealt{torres91}, \citealt{humphreys78}), is also a heavily extinguished ($\mathrm{A}_\mathrm{V}$ = 10.2) object. In combination, these qualities make Cyg OB2 No. 12 excellent for studying extinction caused by dust in the diffuse ISM (\citealt{whittet15}). Consequently, several studies have used the extinction toward Cyg OB2 No. 12 at infrared wavelengths to infer details of the composition of dust in the diffuse ISM (\citealt{rieke74}, \citealt{gillett75}, \citealt{roche84}, \citealt{sandford95}, \citealt{whittet97}, \citealt{bowey98,bowey04}). Similarly, the line of sight toward $\zeta$ Oph, a well-studied bright O9.5 star with $\mathrm{A}_\mathrm{V}$ $\sim$ 1 (\citealt{cardelli89}), is considered representative of diffuse interstellar cloud material (\citealt{savage96}). In this study, we also use Cyg OB2 No. 12 and $\zeta$ Oph as probes of the dust in the interstellar medium. We use stellar atmosphere models and extinction measurements to estimate the optical depth along sight lines toward these objects, then model this optical depth using a suite of various silicate dust opacities to investigate the amorphous dust composition of the diffuse ISM. We also present and compare spectral modelling of the silicate emission from the Trapezium and several pristine spectra of protoplanetary disks. The results of these models are compared and their implications for dust composition are discussed.

\begin{figure*}
\includegraphics[angle=90, width=\textwidth]{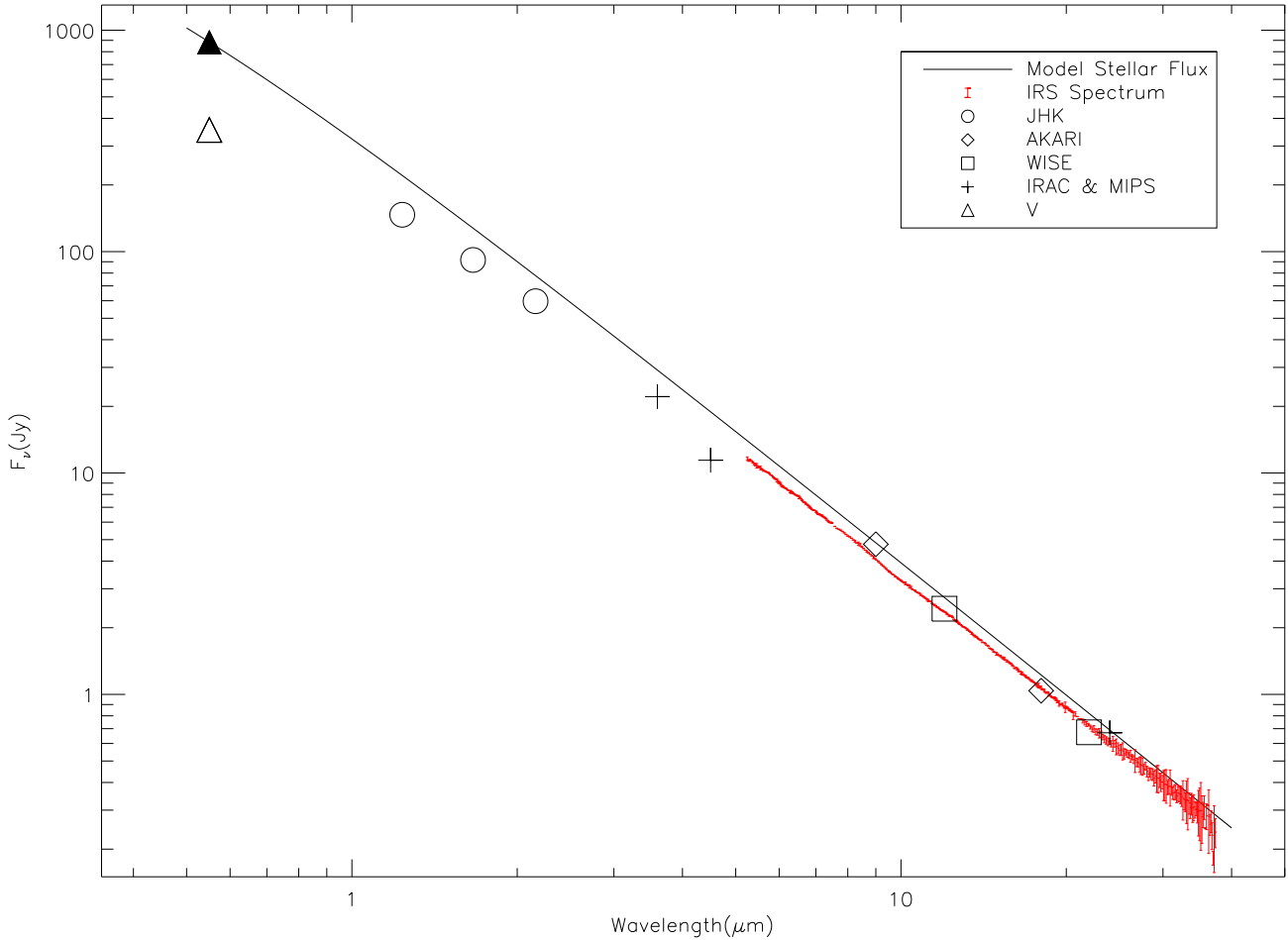} 
\caption{Spectral Energy Distribution for $\zeta$ Oph. The \textit{Spitzer} IRS spectrum is shown in red. The solid line represents the stellar flux, which is a Planck function with an effective temperature of 34,300 K fit to extinction-corrected photometry at the V band. Solid symbols denote photometry corrected for extinction, while open symbols have no correction applied. Photometry is shown from \citealt{herbig68} (V; triangle), \citealt{whittet80} (J, H, K; circle), WISE (3.4, 4.6, 12, and 22$\mu$m; square), IRAC (3.6 and 4.5$\mu$m; plus), MIPS (24$\mu$m; plus), and AKARI (9 and 18$\mu$m; diamond).\label{fig2}}
\end{figure*}

\section{Observations and Data Reduction}
Cyg OB2 No. 12 was observed using the Infrared Spectrograph (IRS) on the \textit{Spitzer Space Telescope} on 2008 August 6 (PI: David Ardila, AOR 27570176). It was observed using the Short-Low (SL) module in all orders (5.2-14$\mu$m) as well as the second order of Long-Low (LL) module (14-21.3$\mu$m). The observations were performed in \textit{Spitzer's} staring mode following a Pointing Calibration and Reference Sensor (PCRS) peak-up observation, which is described in detail by \cite{houck04}. Earlier observations taken on 2004 October 10 (PI: Neal Evans, AOR 9834496) included the first order of LL (19.5-38$\mu$m), and  we include these data as part of the spectral energy distribution (SED) in \autoref{fig1}. The SED presented in \autoref{fig1} also includes photometry from 2MASS, IRAC, MIPS, WISE, and V-band photometry from \cite{caballero14}. $\zeta$ Oph was observed with the IRS with the SL and LL modules on September 10 2008 (PI: David Ardila, AOR 27582976) using \textit{Spitzer's} staring mode following a (PCRS) peak-up observation. The SED of $\zeta$ Oph in \autoref{fig2} presents the infrared spectrum as well as V-band photometry from \cite{herbig68} and near-infrared photometry from \cite{whittet80}.

Observations of the Trapezium presented by \cite{forrest75} at 8-13$\mu$m are combined with a spectrum at 16-38$\mu$m from \cite{forrest76}. The 16-38$\mu$m spectrum corrects for some emission from the Kleinman-Low nebula as described in \cite{forrest76}. This combination of Trapezium spectra was presented by \cite{sargent09} and used to test modelling of silicate emission from protoplanetary disks. We refer the reader to the original papers for observational details.

Observations and data reduction methods for the spectra of five protoplanetary disk spectra of deemed amorphous silicate exemplars in \cite{sargent09} are described in detail in \cite{furlan06}.

The spectra for Cyg OB2 No. 12 and $\zeta$ Oph were downloaded from the Cornell Atlas of \textit{Spitzer} IRS Sources (CASSIS v7; \citealt{lebouteiller11}), using the recommended optimal extraction. We note the IRS spectrum of Cyg OB2 No. 12 exhibits Pfund $\alpha$ and Humphreys $\alpha$ emission lines, likely from a stellar wind. \autoref{fig1} shows the final reduced spectrum for Cyg OB2 No. 12, for which the uncertainties are typically one percent or less in the crucial 10$\mu$m region. Infrared photometry shows good agreement with the IRS spectrum. Good agreement is found between this spectrum and that of Cyg OB2 No. 12 from the Spitzer Atlas of Stellar Spectra (SASS; \citealt{ardila10}). For the spectrum of $\zeta$ Oph, The first order of the SL module was scaled by a factor of 1.03 to match the flux density of adjacent modules. The spectrum exhibits relatively smaller signal-to-noise than that for Cyg OB2 No. 12, especially at longer wavelengths. \autoref{fig2} shows the final reduced spectrum for $\zeta$ Oph.

\section{Modelling}
\subsection{Underlying Flux of Extinguished Stars}
We model the flux from the supergiant star Cyg OB2 No. 12 using a model atmosphere from \cite{Castelli97}.  We used a model atmosphere with an effective temperature of 13,000 K, turbulent velocity of 2 km $\mathrm{s}^{-1}$, and log(g) of 2.5 as used by \cite{bowey04}. 
Using the Kurucz model for the assumed intrinsic colors of Cyg OB2 No. 12 and taking the ratio of $\mathrm{A}_{\mathrm{J}}/\mathrm{A}_{\mathrm{K}_{\mathrm{s}}}$ from the $\mathrm{R}_{\mathrm{V}}$ = 3.1 curve from \cite{mathis90}, we compared the observed J-$\mathrm{K}_{\mathrm{s}}$ color to the intrinsic color to derive an extinction correction for the J band. From the extinction at J we inferred extinction corrections at H and $\mathrm{K}_\mathrm{s}$ using the Mathis extinction curve for $\mathrm{R}_{\mathrm{V}}$ = 3.1. Through this method we estimate $\mathrm{A}_\mathrm{J}$ = 3.3. This extinction estimate corresponds to $\mathrm{A}_\mathrm{V}$ = 11.1, also using the Mathis $\mathrm{R}_{\mathrm{V}}$ = 3.1 extinction curve. This is close to previous estimates, considering the uncertainty in the infrared extinction estimate and its uncertain relation to extinction at visible wavelengths. The Kurucz model atmosphere flux was fit to the extinction-corrected 2MASS $\mathrm{K}_\mathrm{s}$ flux, as shown in \autoref{fig1}. We included a free-free emission component to this model with $\mathrm{F}_{\nu}$ proportional to $\nu^{0.6}$ as in \cite{bowey04}, and tested our model's dependence on this by varying the strength of this free-free emission from 0-20$\%$ of the total flux at 8$\mu$m. The free-free component becomes increasingly important at longer wavelengths since the star's spectrum is close to Rayleigh-Jeans (proportional to $\nu^{2}$) at longer wavelengths. The strength of free-free emission did not significantly impact the resulting comparison between model fits with and without including silicates of intermediate composition in this model.
Photosphere emission from $\zeta$ Oph was approximated as a simple blackbody. A Planck function with an effective temperature of 34,300 K (\citealt{howarth01}) was fit to the photometry in the V-band after correcting for one magnitude of visible extinction. This extinction correction was estimated by comparing the observed B-V color from \cite{moffett79} to intrinsic color of a typical O9.5 V star (\citealt{pecaut13}). The SED for $\zeta$ Oph is shown in \autoref{fig2}.

\capstartfalse
\begin{deluxetable}{lll}
\tabletypesize{\scriptsize}

\tablecaption{Silicate dust in our models\label{tbl-1}}
\tablewidth{0pt}
\tablehead{
\colhead{Name} & \colhead{Abbrev.} & \colhead{Description}
}
\startdata
{\parbox{1.6cm}{Small\\Amorphous\\Polivene}}&SmPoli&{\parbox{4.5cm}{Optical constants for amorphous silicate of Mg$_{1.5}$SiO$_{3.5}$ composition from \cite{jager03}, assuming CDE2 (\citealt{fabian01})}}\\
\\
{\parbox{1.6cm}{Large\\Amorphous\\Polivene}}&LgPoli&{\parbox{4.5cm}{Optical constants for amorphous silicate of Mg$_{1.5}$SiO$_{3.5}$ composition from \cite{jager03}, using the Bruggeman EMT and Mie theory (\citealt{BH1983}) with a vacuum volume fraction of f = 0.6 for porous spherical grains of radius 5$\mu$m}}\\
\\
{\parbox{1.6cm}{Small\\Amorphous\\Olivine}}&SmOl&{\parbox{4.5cm}{Optical constants for amorphous olivine MgFeSiO$_4$ from \cite{dorschner95}, assuming CDE2 (\citealt{fabian01})}}\\
\\
{\parbox{1.6cm}{Large\\Amorphous\\Olivine}}&LgOl&{\parbox{4.5cm}{Optical constants for amorphous olivine MgFeSiO$_4$ from \cite{dorschner95}, using the Bruggeman EMT and Mie theory (\citealt{BH1983}) with a volume fraction of vacuum of f = 0.6 for porous spherical grains of radius 5$\mu$m}}\\
\\
{\parbox{1.6cm}{Small\\Amorphous\\Pyroxene}}&SmPy&{\parbox{4.5cm}{Optical constants for amorphous pyroxene of cosmic composition from \cite{jaeger94}, assuming CDE2 (\citealt{fabian01})}}\\
\\
{\parbox{1.6cm}{Large\\Amorphous\\Pyroxene}}&LgPy&{\parbox{4.5cm}{Optical constants for amorphous pyroxene of cosmic composition from \cite{jaeger94}, using the Bruggeman EMT and Mie theory (\citealt{BH1983}) with a vacuum volume fraction of f = 0.6 for porous spherical grains of radius 5$\mu$m.}}\\
\\
Enstatite&enst&{\parbox{4.5cm}{Opacities for clinoenstatite Mg$_{0.9}$Fe$_{0.1}$SiO$_3$ from \cite{chihara02}}}.\\
\\
Forsterite&forst&{\parbox{4.5cm}{Optical constants for three crystallographic axes of forsterite, Mg$_2$SiO$_4$, from \cite{sogawa06}, assuming tCDE (\citealt{sargent09})}}\\
\\
Silica&silica&{\parbox{4.5cm}{Opacity for annealed silica by \cite{fabian00}}}\\
\enddata
\label{tab1}
\end{deluxetable}
\capstarttrue

\begin{figure*} 
\includegraphics[angle=90, width=\textwidth]{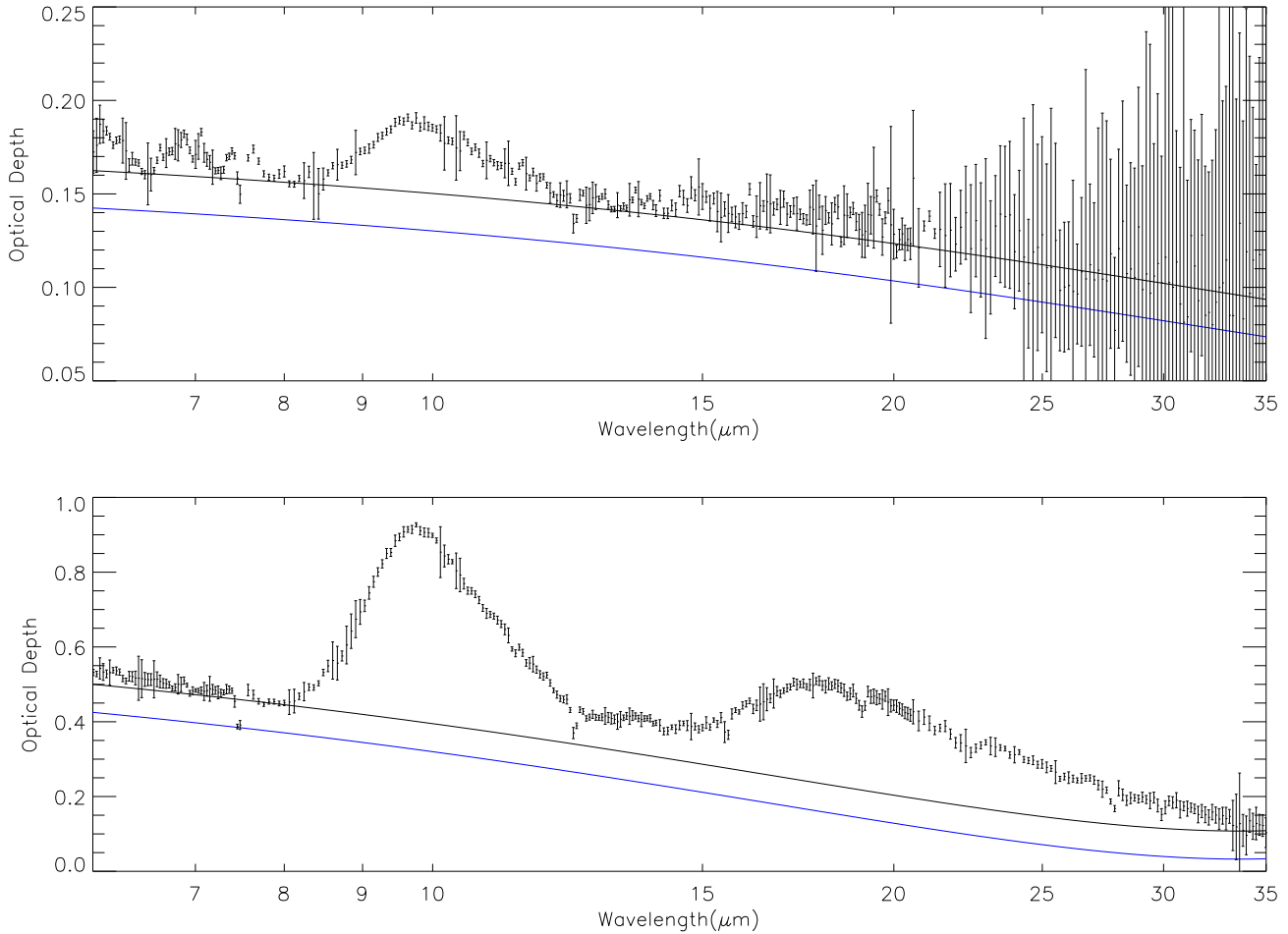} 
\caption{Continuum fit to optical depth toward $\zeta$ Oph (top) and Cyg OB2 No. 12 (bottom). A quadratic fit is made to the lowest points surrounding the silicate features, shown as a black line. This fit is lowered as described in the text, shown as a blue line.\label{fig3}}
\end{figure*}

\subsection{Extinction and Opacity Model}
The ISM dust extinction optical depth is modelled as:
\begin{equation}
\tau(\lambda) = \ln(F_{stellar}(\lambda)/F_{obs}(\lambda))
\end{equation}
where $\tau(\lambda)$ is the optical depth at a given wavelength, and $F_{obs}(\lambda)$ and $F_{stellar}$ are the observed and assumed stellar flux at that wavelength, respectively. The ``stellar'' flux includes the Kurucz model atmosphere flux and any free-free emission assumed. Given this estimate of optical depth, our model finds the best fit to this derived optical depth as the sum of various amorphous and crystalline silicate dust opacities. Our model minimizes $\chi^2$ to find the best fit to the spectrum using the opacities of various silicate dust grains as components. The model components include opacities for amorphous silicate grains of olivine, pyroxene, and intermediate (``polivene'') stoichiometry as well as three crystalline silicate opacities (forsterite, enstatite, and silica). As described in detail in \cite{sargent06, sargent09}, opacities for small grains are calculated assuming grain shapes with a modified continuous distribution of ellipsoid shapes (CDE2, \citealt{fabian01}). Opacities for large amorphous grains are calculated using Bruggeman Effective Medium Theory and Mie Theory (\citealt{BH1983}), using a vacuum volume fraction of f = 0.6 for porous spherical grains of radius 5$\mu$m. We add a ``continuum'' opacity to the model to represent sources of opacity that have little wavelength dependence over the modelled wavelength range. The optical depth as a function of wavelength is modelled using the following equation:
\begin{equation}
\tau(\lambda) = C + \displaystyle\sum_{i} \alpha_i\kappa_i(\lambda)
\end{equation}
where $\tau(\lambda)$ is the optical depth at a given wavelength, C is a constant used as a first order approximation of a ``continuum'' opacity, and each $\kappa_i(\lambda)$ is the opacity of a given dust species, which is scaled by a weighting factor $\alpha_i$. An additional quadratic continuum was estimated by fitting the lowest optical depth values surrounding the silicate features, as seen in \autoref{fig3}. As discussed in \cite{poteet15}, fitting a local continuum can artificially reduce the optical depth contributions from non-continuum sources to zero. To avoid setting the resulting optical depth to zero, but maintain the general shape of the continuum, we slightly reduce the strength of the continuum optical depth (see \autoref{fig3}). The free parameters in this model are the weighting factors and strength of the constant ``continuum'' opacity. As shown in \citealt{min07} and \citealt{bouwman01}, fits to just the $\sim$10$\mu$m feature can be degenerate between properties such as grain shape, size, and iron content. We therefore choose to simultaneously model the $\sim$10$\mu$m and $\sim$20$\mu$m silicate features by fitting a wavelength range of 8-30$\mu$m. We also include a model of only the data from 8-12$\mu$m in order to focus on the $\sim$10$\mu$m feature, avoiding the nearby Pfund $\alpha$ and Humphreys $\alpha$ emission features.

\begin{figure*}
\includegraphics[angle=90, width=\textwidth]{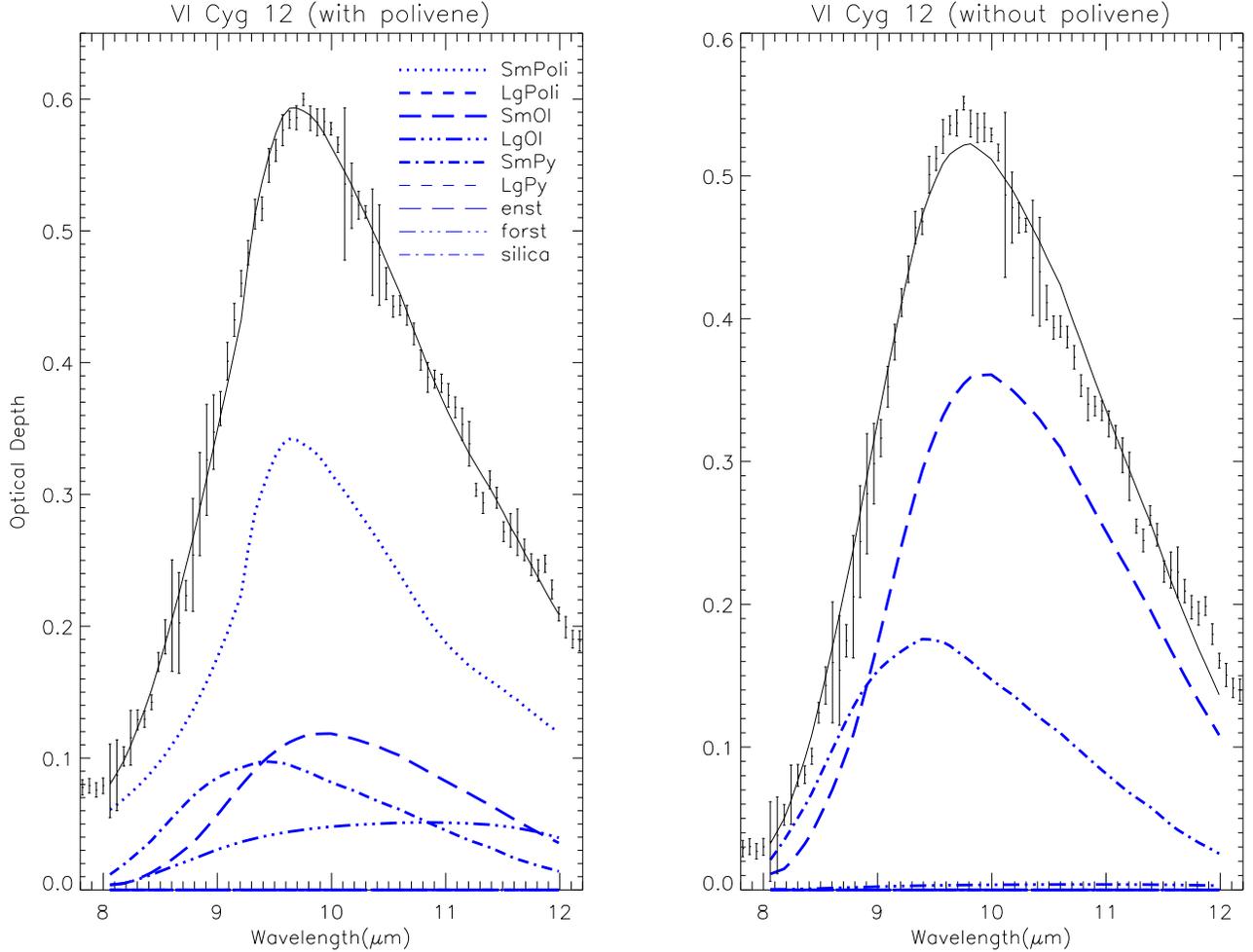} 
\caption{Model fit of optical depth toward Cyg OB2 No. 12, fitting only the silicate feature near 10$\mu$m. The solid line is the total of the individual components shown at the bottom. The left fit includes small and large polivene grain opacities as components, which are disallowed in the right model fit. There are six amorphous silicate components: small and large grain opacities of polivene, olivine, and pyroxene. The three crystalline components are enstatite, forsterite, and silica. The $\chi^2$ per degree of freedom for the fit shown in the continuum-subtracted plots on left and right are 1.8 and 5.6, respectively.\label{fig4}}
\end{figure*}

\subsection{Polivene}
The silicate dust components of our model are listed in \autoref{tab1}. Work by \cite{sargent06, sargent09silica, sargent09}  determined the silicate dust types that best fit the features of protoplanetary disks in star-forming regions like the Taurus-Auriga Molecular Cloud. Since this dust is formed from dust in the ISM, we use the same opacities in our model. In this work, we also include amorphous silicate dust with intermediate stoichiometry (Mg$_{1.5}$SiO$_{3.5}$) to improve fits to infrared spectral features. This dust is neither olivine nor pyroxene, but of intermediate stoichiometry. Though we emphasize that the name does not refer to any (stoichiometric) mineral, we choose to refer to it by the simple shorthand ``polivene.'' \cite{jager03} uses amorphous silicates with intermediate stoichiometry to best represent the dust emissivity of the AGB star TY Dra. Non-stoichiometric grains are found in presolar silicates (see \citealt{nguyen07}, \citealt{nguyen10}, \citealt{vollmer09}, and \citealt{zhao13}). Characterizing sub-micron presolar grains using Auger spectroscopy, \cite{stadermann09} found distinct populations of silicate grains with olivine-like and pyroxene-like stoichiometries, with some grains spanning the gap between the two. Studies by \cite{floss09} and \cite{bose12} find a statistically significant fraction of presolar amorphous silicate grains with intermediate stoichiometry, indicating a distinct population of these intermediate grains. We therefore consider these non-stoichiometric amorphous silicates in our models. 

Using the optical constants for amorphous silicate of Mg$_{1.5}$SiO$_{3.5}$ composition from \cite{jager03}, we calculate opacities for sub-micron grains of polivene composition assuming a CDE2 (\citealt{fabian01}) shape distribution. Like the other large amorphous grains in our model, large polivene grain opacities are derived using Bruggeman EMT and Mie theory (\citealt{BH1983}) with a vacuum volume fraction of f = 0.6 for porous spherical grains of radius 5$\mu$m. The opacity as a function of wavelength for small polivene grains is characterized by the prominent feature peaking at 9.7$\mu$m. Compared to both the opacity of small amorphous olivine and pyroxene in our model, the submicron polivene opacity has a more sharply peaked 9.7$\mu$m feature (see \autoref{fig4}). All three forms of small amorphous silicate opacities in our model have a strong $\sim$10$\mu$m feature that falls off more steeply toward shorter wavelengths, while toward longer wavelengths the peak drops off more gradually. Our model opacities of small olivine and small pyroxene peak at 10.0$\mu$m and 9.4$\mu$m, respectively, while the opacity of small polivene peaks between these two at 9.6$\mu$m. The $\sim$10$\mu$m peak of the small polivene opacity profile for submicron grains is in some ways a combination of that of small amorphous olivine and pyroxene, but its narrowness cannot be reproduced by a linear combination of the opacity profiles of small amorphous olivines and pyroxenes.

The optical constants for the ``polivene'' used in our model are measured from material produced via a sol-gel technique, formed by an aqueous reaction (\citealt{jager03}). These reactions occur at lower temperatures than the regions around AGB stars and red supergiant (RSG) stars where major dust production is thought to occur, as pointed out by \cite{sargent09}. However, the sol-gel method used in earthbound laboratories could potentially create the dust most similar to that in the ISM, created in different environments, such as AGB star outflows. Recently, Koch et al. (2016, in prep) and Sargent et al. (in prep) have also found that non-stoichiometric amorphous silicate opacities improve fits to infrared spectral features of protoplanetary disk spectra in the Orion Nebula Cluster (ONC) and L1641, as well as AGB wind spectra, respectively.

\begin{figure*}
\includegraphics[angle=90, width=\textwidth]{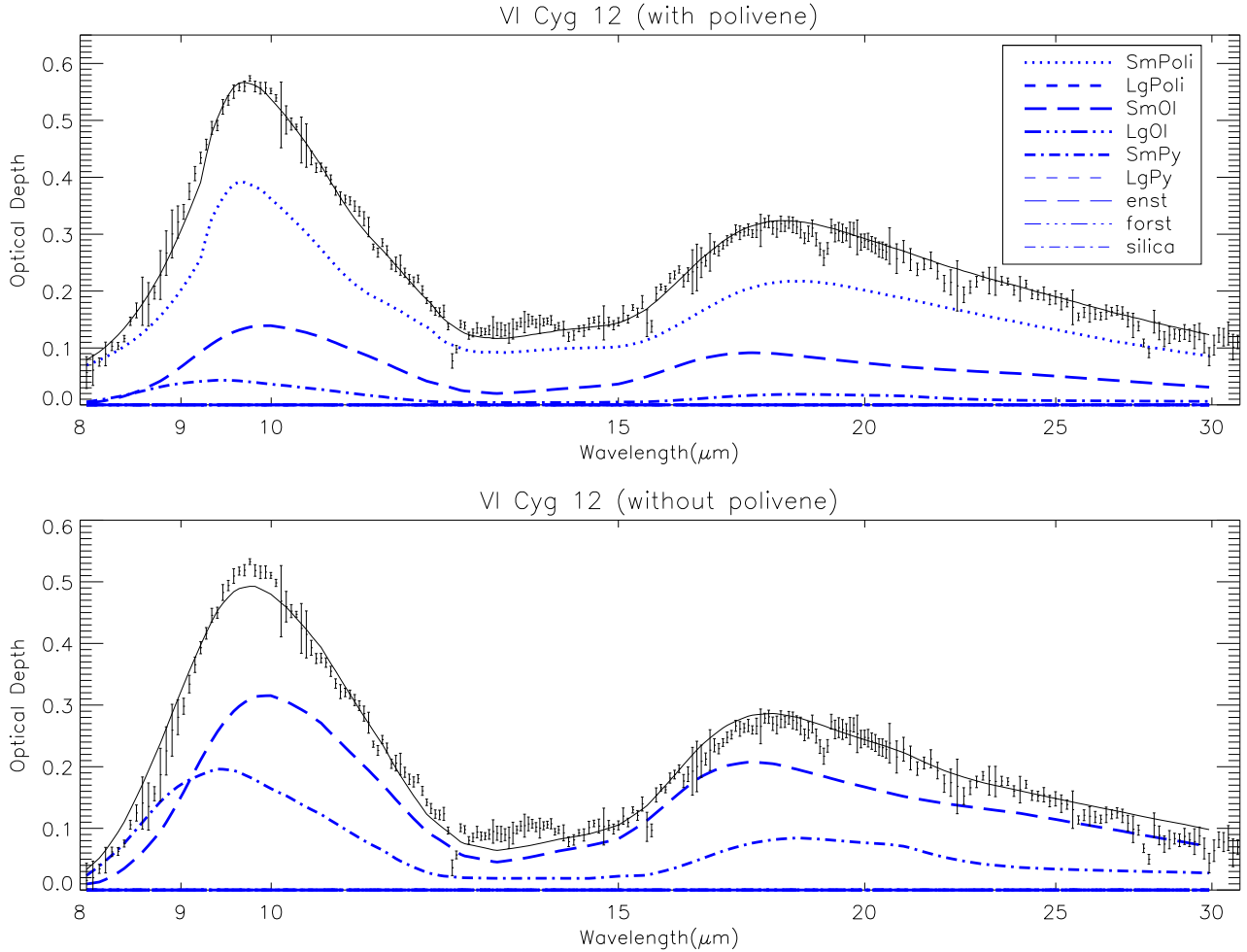} 
\caption{Model fit of optical depth toward Cyg OB2 No. 12, fit from 8-30$\mu$m. The solid line is the total of the individual components shown at the bottom. The top fit includes small and large polivene grain opacities as components, which are disallowed in the bottom model fit. The silicate components are listed in \autoref{tab1}. The $\chi^2$ per degree of freedom for the fit shown in the continuum-subtracted plots on top and bottom are 2.8 and 5.3, respectively.\label{fig5}}
\end{figure*}

\subsection{Two-Temperature Silicate Emission Model}
The spectrum of the Trapezium was modelled assuming that its flux is a combination of thermal emission components from silicate dust grains and simple blackbodies. The model used is described in detail by \cite{sargent09}. This model uses two temperatures which represent a warm component (with warm temperature ``T$_w$'') and cooler component (with cool temperature ``T$_c$'') and models the emission as the sum of emission from two blackbodies of these respective temperatures, added to emission from optically thin dust in the warm (at temperature T$_w$) and cool (T$_c$) regions: 

\begin{subequations}
\begin{eqnarray}
F_\nu(\lambda)^{mod} & = & B_\nu(\lambda, T_c)[\Omega_c + \displaystyle\sum_{i} a_{c,i}\kappa_i(\lambda)]\nonumber\\
& & + B_\nu(\lambda, T_w)[\Omega_w + \displaystyle\sum_{i} a_{w,i}\kappa_i(\lambda)]\nonumber
\end{eqnarray}
\end{subequations}
where $B_\nu(\lambda, T)$ is the blackbody intensity per frequency given by the Planck function at wavelength $\lambda$, $\Omega$ is the solid angle of the corresponding blackbody, and each $a_{i}$ is a mass weight of a component with opacity given by $\kappa_i(\lambda)$ at wavelength $\lambda$. The simple blackbodies are included to represent optically thick continuum emission sources such as those from very large sized grains and amorphous carbon.

We also use this model to fit the spectra of protoplanetary disks. In fact, it is the same two-temperature silicate emission model used in \cite{sargent09} previously used to model protoplanetary disks. For these disks, the model's warm and cool temperature components represent the inner and outer disk, respectively. In this way, the model makes a rough distinction between emission from optically thin dust in the inner and outer regions of the disk. 

\section{Results}

\subsection{Interstellar Polivene}

The modelling of optical depth along lines of sight toward Cyg OB2 No. 12 and $\zeta$ Oph using the infrared opacities of a suite of silicate dust opacities reveals evidence for interstellar polivene. The optical depth modelling in both cases is improved by the inclusion of polivene. The models used in each figure are identical, except for the inclusion of amorphous polivene grains in the top models, which improves the model fit. 

\begin{figure*} 
\includegraphics[angle=90, width=\textwidth]{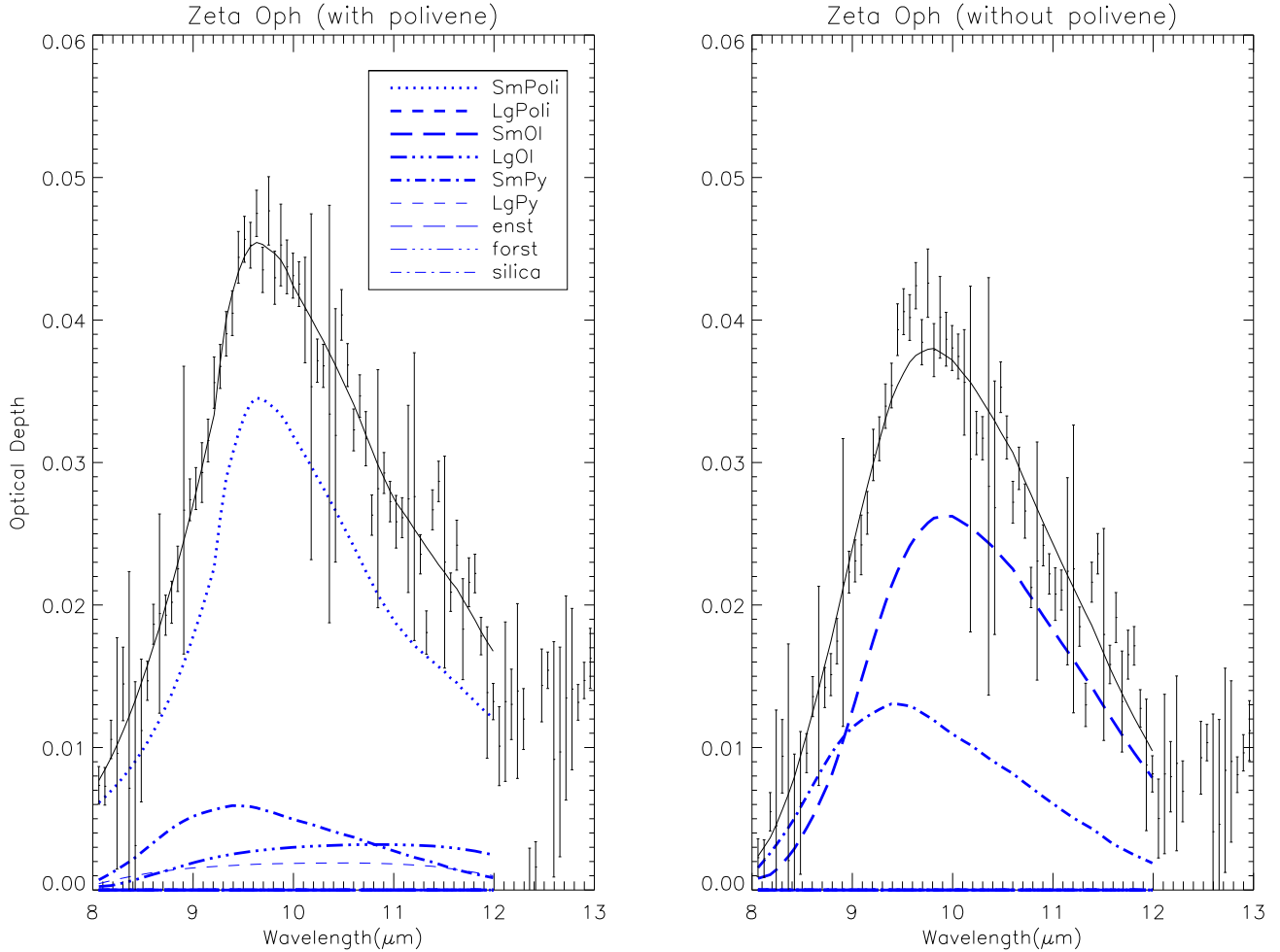} 
\caption{Model fit of optical depth toward $\zeta$ Oph, fitting only the silicate feature near 10$\mu$m. The solid line is the total of the individual components shown at the bottom. The left fit includes small and large polivene grain opacities as components, which are disallowed in the right model fit. The silicate components are listed in \autoref{tab1}. The $\chi^2$ per degree of freedom for the fit shown in the continuum-subtracted plots on left and right are 1.7 and 3.1, respectively.\label{fig6}}
\end{figure*}

\begin{figure*} 
\includegraphics[angle=90, width=\textwidth]{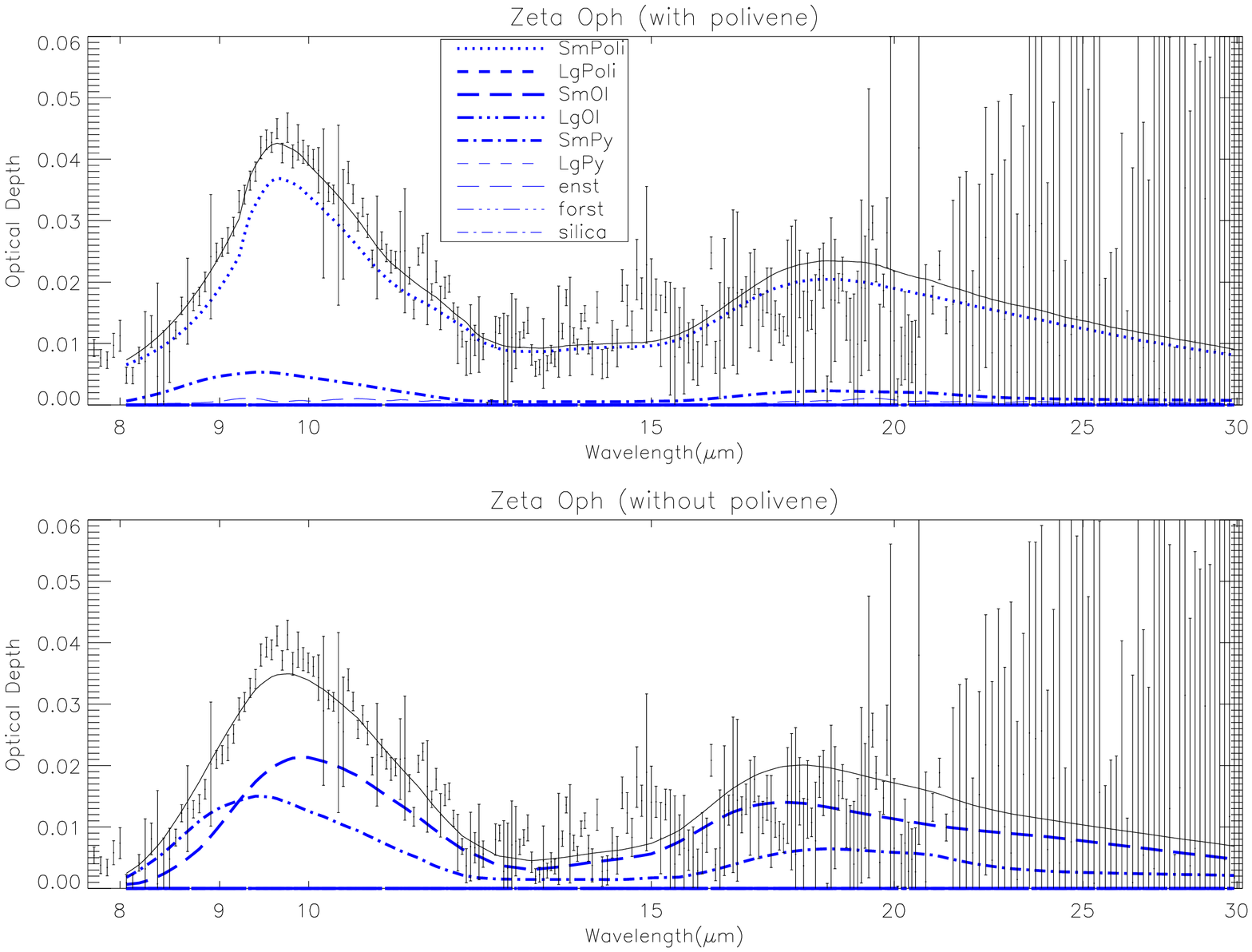} 
\caption{Model fit of optical depth toward $\zeta$ Oph, fit from 8-30$\mu$m. The solid line is the total of the individual components shown at the bottom. The top fit includes small and large polivene grain opacities as components, which are disallowed in the bottom model fit. The silicate components are listed in \autoref{tab1}. The $\chi^2$ per degree of freedom for the fit shown in the continuum-subtracted plots on top and bottom are 2.8 and 3.6, respectively.\label{fig7}}
\end{figure*}

In both Cyg OB2 No. 12 and $\zeta$ Oph, when polivene is included as a component in the modelling, the contribution to the optical depth is dominated by small amorphous polivene. Including polivene improves the model fit for Cyg OB2 No. 12 (see \autoref{fig5}) from a reduced $\chi^2$ value of 5.3 (without polivene) to a reduced $\chi^2$ of 2.8 (with polivene). When just focusing between 8-12$\mu$m, these respective reduced $\chi^2$ values are 5.6 and 1.8 (see \autoref{fig4}). An improvement is also seen in $\zeta$ Oph's model fit improvement from a reduced $\chi^2$ value of 3.1 (without polivene) to a reduced $\chi^2$ of 1.7 (with polivene) when fitting from 8-12$\mu$m to focus on the $\sim$9.7$\mu$m feature (see \autoref{fig6}). Modelling $\zeta$ Oph over a larger wavelength range of 8-30$\mu$m yields an improvement in reduced $\chi^2$ from 3.6 to 2.8 after polivene is included (see \autoref{fig7}). For both objects, the optical depth near 10$\mu$m is characterized by a steep increase between 9$\mu$m and 9.5$\mu$m, and a more gradual decline beyond the peak at $\sim$9.7$\mu$m. This general shape is reproduced well by the opacity of polivene, leading to a significant improvement in the model fit.

\begin{figure*}  
\includegraphics[angle=90, width=\textwidth]{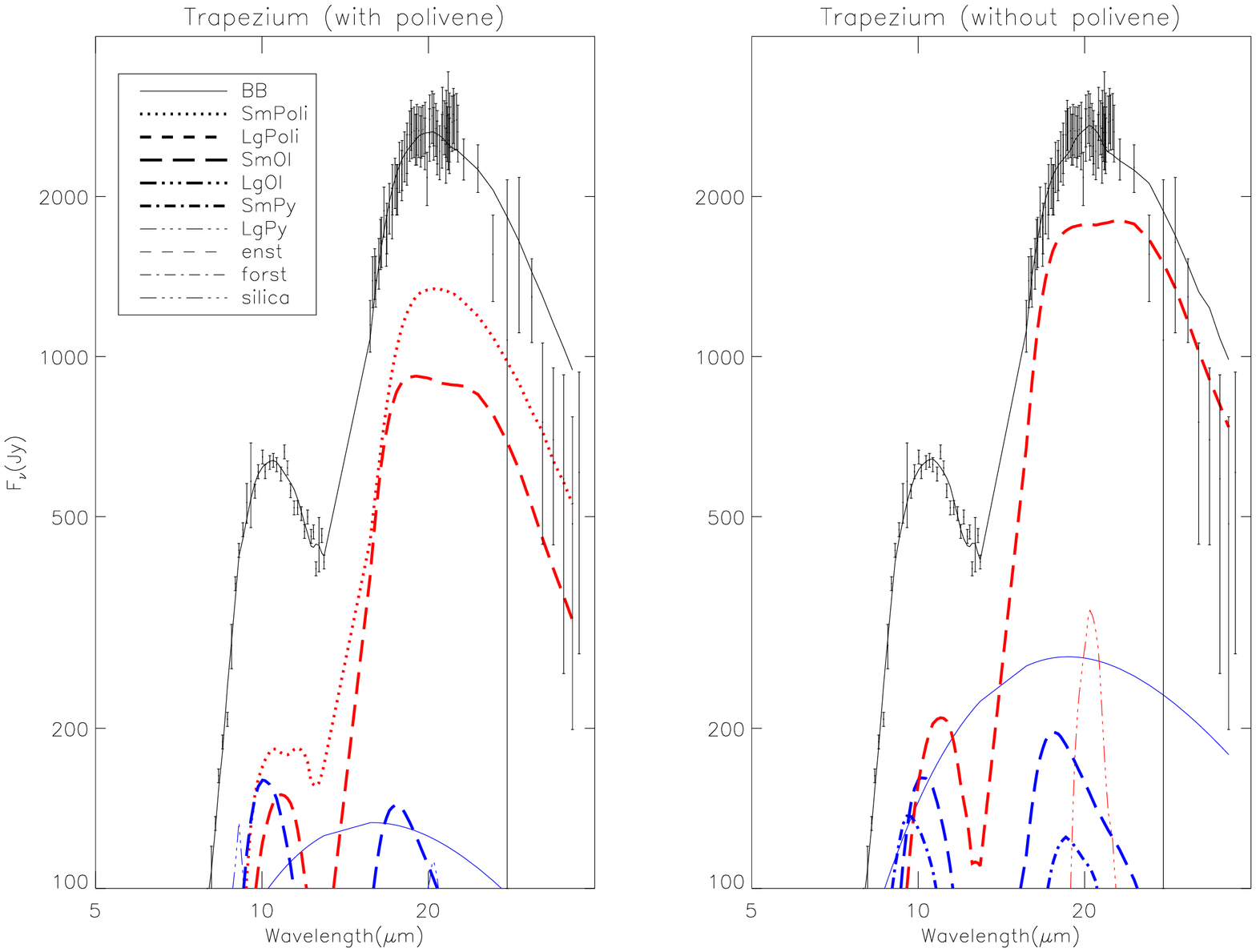} 
\caption{Model fit of silicate emission from the Trapezium. The solid line is the total of the individual components shown at the bottom. The left fit includes small and large polivene grain opacities as components, which are disallowed in the right model fit. The color indicates components at either the warm (blue) or cool (red) temperature in this model. Simple blackbodies represented by Planck functions at each of the two temperatures are included as well as warm and cool emission from the dust species listed in the previous figure. The $\chi^2$ per degree of freedom for the fit shown in the plots on left and right are 1.48 and 1.60, respectively.\label{fig8}}
\end{figure*}

The two-temperature silicate emission model was used to analyze the spectrum of the Trapezium. \autoref{fig8} shows the best fit to the Trapezium spectrum, for both the case where polivene is included and when it is not. The $\chi^2$ per degree of freedom is improved in our model when polivene is included, but only from a value of 1.60 to 1.48, which is not considered particularly significant. When allowing polivene, the model's cool dust component is dominated by emission from small-sized polivene grains. Best fit model parameters are listed in \autoref{tab3}.

\subsection{Polivene in Protoplanetary Disks}
With the strong evidence for polivene in the ISM, and since dust in the ISM makes up the seed material for protoplanetary disks, it is important to include polivene in studies of protoplanetary disks. Given the evidence for polivene in the ISM, we investigate the disks in the Taurus-Auriga molecular cloud, a well-studied region of star formation. The dust in the protoplanetary disks of Taurus has been characterized by several studies (e.g. \citealt{furlan06, furlan09, furlan11}; \citealt{watson09}; \citealt{sargent06, sargent09}). \cite{sargent09} identifies several ``amorphous silicate exemplars'' in Taurus, which are protoplanetary disks whose infrared spectra show only the very smooth and broad features indicative of amorphous silicate dust. Our two-temperature model characterizes the dust in five of these amorphous silicate exemplars. As shown in \autoref{tab2}, fits to the infrared spectrum of these amorphous exemplars are not significantly improved by adding polivene opacities as a component in the model. Most of these fits were only improved so slightly that the $\chi^2$ per degree of freedom decreased by 0.01-0.05. The exception to this was LkCa 15, as the fit improved in reduced $\chi^2$ by 0.31. However, this is mostly due to the poor general fit to the spectrum in both cases at longer wavelengths, for which the model also adds a sizable forsterite component that has been deemed an artifact (see \citealt{sargent09}). In addition, the model results had very small, if any, polivene masses in their very slightly improved best fits. Taurus protoplanetary disk spectra show no strong evidence for non-stoichiometric amorphous dust emission. 

\capstartfalse
\begin{deluxetable}{lcc}
\tabletypesize{\scriptsize}
\tablecaption{Modelled Fit $\chi^2$  per d.o.f. for disks in Taurus\label{tbl-2}}
\tablewidth{0pt}
\tablehead{
\colhead{Name} & \colhead{Without Polivene} & \colhead{With Polivene}
}
\startdata
CoKu Tau 4&2.42&2.37\\
DM Tau&4.82&4.81\\
FM Tau&2.43&2.40\\
GM Aur&5.04&5.03\\
LkCa15&7.78&7.47\\
\enddata
\label{tab2}
\end{deluxetable}
\capstarttrue

\section{Discussion}
Modelling the dust composition of the ISM and protoplanetary disks using opacities for amorphous dust of intermediate stoichiometry between that of olivine and pyroxene (``polivene'') gives new insight into dust formation and processing in astrophysical contexts. Using opacities of various silicate dust species, we have modelled the optical depth of the intervening ISM material along lines of sight toward Cyg OB2 No. 12 and $\zeta$ Oph. With this model, we find strong evidence for a population of interstellar polivene. This suggests an efficient mechanism for producing sub-micron sized polivene dust grains, presumably from red giant branch stars (RGB) and asymptotic giant branch (AGB) stars. More work is needed to determine if polivene is ubiquitous in the ISM or if it is found only in certain regions. Evidence for polivene in the ISM is consistent with meteoritic evidence of non-stoichiometric amorphous compositions found in pre-solar grains. In the formation of the solar system, these presolar grains were protected inside meteorites and remained relatively unprocessed throughout the last 4.567 Gyr. These non-stoichiometric amorphous grains found in primitive meteorites represent interstellar dust that populated a molecular cloud from which the solar system formed. Our model's result of interstellar dust toward Cyg OB2 No. 12 and $\zeta$ Oph supports the hypothesis based on meteoritic evidence that non-stoichiometric amorphous silicate dust is present in the diffuse ISM.

The best fits to the modelled optical depth toward Cyg OB2 No. 12 and $\zeta$ Oph exhibit general agreement that the dust in the ISM responsible for the broad absorption features is mostly amorphous, small silicate grains. The best fits include only amorphous dust as components. The strength of the large amorphous grain components in our model are much weaker than the dominating features from smaller (submicron) grains. The small amorphous silicates used in our modelling have equal Fe and Mg content, with the exception of polivene, which includes no Fe. The ISM toward $\zeta$ Oph is also found to consist of small amorphous silicates with Fe/Mg ratios $\lesssim$ 1 by \cite{poteet15}, who use amorphous silicates of olivine and pyroxene stoichiometry with varying Fe/Mg ratios in their modelling and find a lack of the most Fe-rich silicates in their best fits. However, as noted by \cite{sargent09} and \cite{poteet15}, there exists significant degeneracy between Fe-rich and Fe-poor silicates in these models. This degeneracy inhibits the ability these models to draw conclusions about the Fe content of interstellar silicate dust. When polivene is not included in our models, small amorphous grains of olivine stoichiometry dominate the composition of the ISM toward $\zeta$ Oph, rather than pyroxene-like grains. However, we note that \cite{sargent09} shows significant degeneracy between amorphous grains of olivine and pyroxene compositions. \cite{poteet15} models the IRS spectrum of $\zeta$ Oph and also finds that amorphous grains of principally olivine-like composition provide the best model fit to the spectrum. Just as in our best fit of the ISM toward $\zeta$ Oph without including polivene, the model fit in \cite{poteet15} has a slightly broader peak near 9.7$\mu$m that is shifted to a slightly longer wavelength than that of the data. The addition of amorphous silicate grains of intermediate stoichiometry adds a component to these interstellar dust models that improves fits to the prominent 9.7$\mu$m feature.

Protoplanetary disks, which form inside molecular clouds of gas and dust, represent another stage of dust processing in the journey toward planet formation. If polivene is present in the diffuse ISM, then we expect that the molecular clouds from which protoplanetary disks form also harbor polivene. However, our modelling of amorphous silicate exemplar disks in Taurus lack any significant evidence of polivene. One possible explanation for this discrepancy would be significant non-uniformity of polivene populations in the ISM. This explanation would mean that the Taurus-Auriga molecular cloud formed from interstellar material that was much less rich in polivene than along a sightline toward Cyg OB2 No. 12 and $\zeta$ Oph. Additional studies would be required to determine possible polivene presence along lines of sight other than toward Cyg OB2 No. 12 and $\zeta$ Oph. Evidence of this significant non-uniformity would suggest dust production methods that lead to regions of high and low concentrations of polivene in the interstellar medium. 

\capstartfalse
\begin{deluxetable*}{lccccccccccc}
\tabletypesize{\scriptsize}
\tablecaption{Modelled Fit $\chi^2$  per d.o.f. for various sightlines\label{tbl-3}}
\tablehead{
\colhead{Object Name} & \colhead{Polivene in model?} & \colhead{SmPoli} & \colhead{LgPoli} & \colhead{SmOl} & \colhead{LgOl} & \colhead{SmPy} & \colhead{LgPy} & \colhead{enst} & \colhead{forst} & \colhead{silica} & \colhead{$\chi^2$/d.o.f.}  
}
\startdata
Cyg OB2 No. 12&Yes&74.4&0&20.3&0&5.3&0&0&0&0&2.8\\
Cyg OB2 No. 12&No&-&-&66&0&34&0&0&0&0&5.3\\
$\zeta$ Oph&Yes&90.7&0&0&0&9.3&0&0&0&0&2.8\\
$\zeta$ Oph&No&-&-&63.2&0&36.8&0&0&0&0&3.6\\
Trapezium&Yes&0-66&0&63-34&0&18-0&0&0&0&20-0&1.5\\
Trapezium&No&-&-&51-95&0&41-0&0&0&0&8-5&1.6\\
\enddata
\tablecomments{Best model fit parameters for sightlines toward Cyg OB2 No. 12, $\zeta$ Oph, and the Trapezium. Each dust parameter (see \autoref{tab1} for abbreviations and details) is listed as a mass percentage, excluding continuum emission. The values listed here are for the model fits spanning a  wavelength range from 8-30$\mu$m. In the case of the Trapezium, the two values listed for each dust type correspond to the mass percentages of the modelled warm and cool regions, respectively.}
\label{tab3}
\end{deluxetable*}
\capstarttrue

Our results for disks in Taurus also differ from recent results regarding protoplanetary dust in the disks of the ONC and L1641. Work by Koch et al. (2016, in prep) finds that including polivene in models of disk emission significantly improves fits to the infrared spectra of protoplanetary disks in the ONC and L1641. Koch et al. finds 57 of 76 disks showed evidence for polivene in the ONC, and 53 of 72 in L1641. This would be further demonstration of potential non-uniformity in polivene distribution. Dust grain processing could also account for the evidence of polivene in Orion disks and lack of polivene emission in the disks of the Taurus-Auriga molecular cloud complex. Hertzsprung-Russell diagrams show that T-Tauri stars in Taurus are significantly older than those in Orion (\citealt{kenyon95}, \citealt{hillenbrand97}, \citealt{luhman09}, \citealt{dario10}). Comparison to models (e.g. \citealt{siess00}) indicates a median age difference of $\sim$1 Myr. Since the lifetimes of Class II objects are likely underestimated (\citealt{bell13}), this median age difference may be even larger. If the polivene grains present in the ISM undergo processing to produce amorphous grains of olivine and pyroxene stoichiometry, that processing could explain the lack of polivene in the older, more processed Taurus disks. According to this hypothesis, protoplanetary disks in the youngest nearby regions such as NGC 1333, Serpens, and the core of $\rho$ Ophiuchus should be more likely to exhibit evidence for polivene than the disks in older, more processed regions like  Scorpius-Centaurus, Chamaeleon, or the off-core region of $\rho$ Ophiuchus. 

Future studies could test these hypotheses, filling in our understanding of key phases of dust evolution. The \textit{Spitzer} mission obtained spectra for circa 2000 nearby protoplanetary disks. Further analysis of the dust composition within these disks and how the compositions correspond to age, locale, etc. could test our understanding of dust processing and non-uniformity. Studying details of dust composition in these key phases informs our understanding of dust grains' long journey, from the production mechanisms in AGB and RGB winds, travelling through the diffuse ISM, gathering into molecular clouds and collapsing into the protoplanetary disks where they are processed and coalesce to finally form planets.

\section{Conclusions}
We analyzed the composition of interstellar dust along lines of sight toward Cyg OB2 No. 12 and $\zeta$ Oph. First, the optical depth toward these objects was estimated using the infrared spectrum, photometry and stellar atmosphere models. Modelling the optical depth toward Cyg OB2 No. 12 and $\zeta$ Oph at infrared wavelengths as a sum of silicate opacities and a continuum, we find that the optical depth of extinguishing material toward the object Cyg OB2 No. 12 and $\zeta$ Oph exhibit strong features indicative of amorphous silicate dust with intermediate stoichiometry between olivine and pyroxene. We conclude that a significant ``polivene'' dust population exists in the diffuse interstellar medium toward Cyg OB2 No. 12 and $\zeta$ Oph. 

The infrared spectrum of the Trapezium was modelled as a combination of a cool and a warm component of thermal emission from simple blackbodies and silicate dust species. The best model fit to the Trapezium spectrum is only slightly improved by the inclusion of polivene. We conclude that our model shows the Trapezium spectrum to be consistent with an interstellar polivene dust population.

The dust compositions of five protoplanetary disks in Taurus whose infrared spectra are dominated by amorphous silicates were investigated. We studied these spectra by modelling the disk emission as a warm inner region and a cooler outer region. In each region a simple blackbody contribution was added to the emission from each of the nine dust species, and the model found a best fit to the infrared spectrum using these components. Our study found no significant evidence for polivene dust grains in the emission features of these disks of Taurus. We conclude that the dust in Taurus must have either coalesced from a region of the ISM that was significantly less rich in polivene than along the sight lines toward Cyg OB2 No. 12 and $\zeta$ Oph, or that it has been largely processed away in the somewhat older Taurus disks.

\acknowledgments

We thank the referee for a careful review and helpful comments that led us to improve this paper. This work is based in part on observations made with the Spitzer Space Telescope, obtained from the NASA/IPAC Infrared Science Archive, both of which are operated by the Jet Propulsion Laboratory, California Institute of Technology under a contract with the National Aeronautics and Space Administration. This work was supported by NASA under grant number NNX11AD27G. BAS acknowledges funding from NASA ADAP grants NNX13AD54G and NNX15AF15G.



\bibliography{references}

\end{document}